# Exploring Landscape, renormgroup quantization

Maxim Budaev

11.01.2008

## Abstract

In this paper the Landscape potential is considered as an environment for system: trajectory-environment (TE). The trajectory is generating a measure on the landscape. The entropy of this dynamic measure is a power factor for trajectory. This dynamics leads to a memory appearance and produces a non-singular measure practically independently from initial conditions. It is shown that measure is dual to the metrics and its evolution may be considered as the landscape deformations: production-destruction of attractors-vacua. It seems like the renormalization process with phase transitions: renormgroup quantization. It is shown what the entropy of this global landscape measure acts on the trajectory alike a dark energy.

---

Email: max.budaev@yahoo.com

# 1 Introduction

The Landscape paradigm [1][2] demands a dynamics for self exploring. This dynamics must define the lifetimes for the metastable vacua and a non-causal interaction between them. Discrete, dense spectrum of a landscape potential requires a dynamics, which has a jump regime alternating by a perturbative one. For these purposes, one usually uses an ensemble: the ensemble of initial conditions [7][8][9][10][11][12].

In this paper we shall consider a constructive approach to these purposes. Namely, we shall consider the trajectory of exploratory process as a sequence of measurements-decoherenses, which produces the configuration space in a form of measure. We shall see what this dynamic measure is dual to the geometry of landscape potential: to a metrics in Kähler sense. In considered model (TE) the entropy of this measure is a real conformal dynamic factor for a metrics on configuration space. This dynamics leads to a memory appearance. The memory is remarkably emerges on large scale (on scale of attractors' (vacua) dynamics) as a topological (non-perturbative) convergence to a global potential's flatness. This convergence does not lead to stationarity. There are enabled the large fluctuations which uplifts the image of trajectory in the landscape energy spectrum. Next "relaxation" is a sequence of metastable universes with tendency to supersymmetry, and so on.

In the renormalization domain, the measure produces an effective (renormalized) environment. We shall see that the dynamics on landscape spectrum may be considered as a renormgroup quantization.

The paper is organized as follows.
§2- §7 partly review of [3].
In §2 we define the attractor (metastable event) and characterize it in terms convenient for following estimations.
In §3 we prove the bifurcation relation (duality condition)–the important result for local dynamics.
In §4 we estimate the lifetime of attractor and hence show what attractor is not exactly attractor; this object has several phases of evolution including a destruction phase. During the lifetime estimation we shall see these phases in detail. We define important value-the entropy production-the bifurcation production.
In §5 the scale of attractors' dynamics is considered. We shall see that sequence of attractors' lifetimes is correlated by means of memory.
In §6 we shall consider the entropy evolution in more detail and estimate a fluctuation time and a relaxation time.
In §7 the cycles in a sequence of metastable vacua are considered.
In §8 we consider the phase of attractor's destruction in the dark energy context.
§9 is devoted to discussion.

## 2 Dynamics

The gradient flow of a landscape potential has a vast set of the critical submanifolds of various dimensions and geometry: the attractors-vacua that draws the trajectories. Having added a small stochastic term to gradient-like system, it is possible to abstract from unstable, critical submanifolds. As a result the system of the Langeven type is appearing:

$$\Delta\varphi = -\lambda(t)\cdot\nabla_{\varphi}U(\varphi)\cdot\Delta t + \Delta\eta$$

$t \in \mathbb{N}$ -discrete time: $\Delta t = 1$,

$\varphi \subset M$ - configuration space-outcome space,

$\varphi(t)$ - culminating point of trajectory.

In future, for simplicity, we shall name $\{\varphi_t\}$ and $\varphi(t)$ by the "trajectory".

$U \in C^{\infty}(M)$ - environment-landscape potential,

$\eta(t), \langle\eta(t)\rangle = 0$ - stochastic process-quantum fluctuations. (1)

We shall assume, for simplicity, the noise is uniform distributed in a small bounded interval $(-\eta, \eta)$. The noise is important, first of all, as a "destruction mechanism" for unstable, critical submanifolds of gradient flow.

$$\lambda = \lambda(t) \in \mathbb{R} \text{ -bifurcation (renormalization) parameter-power factor.}$$

In fact, it is the metric factor $\lambda \equiv \lambda(t)\cdot\delta_{ij}$, is defined by means of measure. For definition of the measure is generated by a discrete trajectory embedded in a configuration space, that may have arbitrary dimension, we consider some partition $\xi$ of the configuration space on events (coarse-grain).

We assume the average diameter of partition's cells much more then noise amplitude-$\eta$.

Let's define the dynamic frequency measure $\mu$ on this partition: Let at the time $T$ the trajectory has spent the time- $n_i(T)$ in a partition's cell $\xi_i$. Then we define:

$$\mu_i(T) \equiv \frac{n_i(T)}{T}.$$

Obviously,

$$\sum_{i \in V} \mu_i(T) = 1 \text{ for all } T.$$

Here $V(T) \subset \xi$ -set of "visited cells" at the time $T$.

Thus, $\mu \in Sym(\mathbb{Q}^{|V|})$.

Now define $\lambda$.
**Definition 1**
Entropy of individual trajectory is:

$$h(t) \equiv h(\xi, t) = -\sum_{i \in V} \mu_i(t) \cdot \log \mu_i(t).$$

Formally, it is the entropy of measure $\mu$ on partition $\xi$.
Let

$$|V| \equiv |V(T)| \in \mathbb{N}$$ - is capacity of visited cells set.

At last, we put by definition:

$$\lambda(T) \equiv \lambda_0 \cdot \left(\log|V| - h(T)\right) \qquad (2)$$

Here $\lambda_0$ -positive constant, dimensional "inverse mass".

Note, that the bifurcational parameter $\lambda$ is a nonincreasing function of formal entropy.
$\lambda$ can be considered also as a dynamic variable, measuring a deviation of the current distribution from uniform one. In this sense the bifurcational parameter is an information factor. On the other hand, it is a momentum scale that is a power factor.
In the dual, $\lambda$-functional is a conformal factor for metrics, which depends on the trajectory's "mass distribution" in configuration space (2). Thus, the density's fluctuations are expand/compress the configuration space. In this sense $\lambda$ is a geometry factor (curvature).

## 3 Bifurcational relation

The bifurcational relation (duality condition) is an inequality establishing communication between the information and the local geometry (energy) of a landscape potential during the moment of transition between attractors-vacua.
First of all, we characterize attractors in terms convenient for further estimations. For this purpose we shall define attraction area i.e. low-energy limit. First, we consider the elementary attractor-local minimum. In some vicinity $O$ of a local minimum the landscape potential can be presented in a quadratic form with some accuracy. After reduction to main axes (proper basis) we obtain:

$$U(\varphi) = U(\varphi_0) + \sum_{i=1}^{n} \omega_i \cdot (\varphi_i - \varphi_0)^2 + o((\varphi - \varphi_0)^2).$$

Here $\varphi - \varphi_0$ -local coordinates corresponding to vacuum $\varphi_0$. These coordinates corresponds to fundamental fields of a local Lagrange theory. $\omega_i \equiv \omega_{ii} \geq 0$ - local environment's curvature tensor that encoding the symmetry breaking hierarchy ( $\omega_{ij}$ - matrix of couplings and masses of local quadratic theory. Here we shall name this the *moduli space*). The equality sign corresponds to possible degeneration in some directions in case of a non-trivial local minimum (critical submanifold). Let's emphasize what the different local minima can have various codimensions. See [3] for more detail.

**Remark**
The culminating point of the trajectory in vicinity of a critical submanifold may be considered as a Feynman's primitive tree diagram (worldsheet). Then the trajectory is an integral process of gluing the actual diagram to a horizon of multiverse state with bulk (history) dependent couplings (effective environment).

In the further, for estimations, we shall characterize the attractor by scalar:

$$\varpi = \frac{1}{N}\sum_{i=1}^{N}\omega_i$$ - "local average curvature" or "energy" of attractor.

$N$ - attractor's codimension.
In estimations we shall use the partition associated with a set of all attractors $A$:

$$\forall a \in A, \exists \xi_a \in \xi : O(a) \subset \xi_a,$$

and $A \leftrightarrow \xi$ -is a biunique correspondence. It is possible to tell, what the cell corresponds to the moduli.
Let the energy of attractor, which contains in cell $\xi_i$, is $\varpi_i$. We shall name set: $\{\varpi_i\}$ the *power spectrum of landscape*.
Thus we identify the partition and set of attractors and characterize them by average curvatures:

$$\{A_i\} \leftrightarrow \{\xi_i\} \leftrightarrow \{\varpi_i\}.$$

Further the symbol $\omega_i$ will designate event, attractor or attractor's energy (curvature), depending on context.

**Proposition 1**
The bifurcation relation (duality condition) is:

$$\lambda \cdot \omega > 1.\qquad(3)$$

**Proof:**
The qualitative picture of system's behavior in a vicinity of a local minimum is such:
After the trajectory gets in basin of attraction at the time
$T : \varphi(T) \in O(\varphi_0), \varphi_0 \in A$, it quickly rolls down on a bottom of hole:

$$\varphi \xrightarrow{t} \varphi_0,$$

$$t \sim O(1).$$

Then the trajectory oscillates in a vicinity of a bottom with amplitude order of

$$|\Delta\varphi| \simeq \lambda(T) \cdot 2 \cdot \varpi \cdot \eta_0.$$

Further, $\lambda$ - function varies in complicated manner until its magnitude does not become sufficient for performance of condition:

$$\lambda(T) \cdot 2 \cdot \varpi \cdot |\varphi - \varphi_0| > 2 \cdot |\varphi - \varphi_0|,$$

or

$$\lambda(T) > \varpi^{-1}.$$

Differently, the drift factor should change (increase) up to such magnitude that an each next iteration moves away a trajectory from "bottom" of a potential hole.

**Remark**

The fact, what $\lambda$-function grows in this case, will be discussed in detail at estimation of the attractor's lifetime. It is also visible from qualitative picture: localization of the trajectory in attractor (measure concentration effect) is accompanying by entropy decreasing (increasing of conditional entropy) and by increasing of $\lambda$ - function ([2](#)).

After performance of an inequality ([3](#)) the trajectory very fast leaves the attractor. "Very fast" means:

$$\log \delta(t) \sim t^2,$$

$\delta(t)$-scale of the trajectory deviation from a bottom of potential hole.

Here $t$ - local time.

Really

$$\partial_t \delta \sim \lambda(t) \cdot \delta,$$

and, as it will be visible (a curve into black rectangle on Figure [1](#)), locally:

$$\partial_t \lambda \approx const > 0,$$

$$\lambda - \omega^{-1} \sim t$$

We see, that the trajectory is "blow-up tunneling" (superinflation) through a potential barrier. It states a necessary estimation and permits finish the proposition proof.•

The trajectory in space of potentials:

$$\varpi(t) \equiv \lambda_0^{-1} \cdot \lambda(t) \cdot \omega \qquad (4)$$

is required for us. We shall name this by *effective curvature of potential*. It is an illustration of a trajectory-landscape duality. In the renormalization sense this defines a renormgroup integral for effective couplings and masses. This is analogous to the "effective" central charge in the conformal field theory. If the space of effective couplings it is chosen as a landscape potential curvature.

Some important remarks it is appropriate mention here.

1. The blow-up inflation is allowing us to have no interest in the structure of a potential barrier (width, height, etc.). We shall assume that the trajectory have an "instant jump" into next vacuum after performance the bifurcation relation ([3](#)).
2. If the trajectory gets in some attraction area (vacuum) and the bifurcation relation ([3](#)) is attained, the attractor's lifetime will be insignificant: $\tau \sim 1$. It is possible to tell, that the trajectory still "does not notice" this attractor. Or, in psychophysical terminology, the subject (trajectory) still remembers such or a "more strong" geometry, and it does not the big attention (interest) for actual event. Take into account these reasons and that the individual trajectory instead of ensemble is considered, in the further we shall use concept of event instead of a state. Here it is pertinent to specify the concept of metastability in our context.

**Definition 2**

The event is called *metastable or singular* on condition that:

$$\varpi(t) \leq \lambda_0, \qquad (5)$$

and *coherent* otherwise.

The metastable events quantitatively differ from coherent they by much greater time of life. The concept coherent is accepted in the linear dynamics. In our approach it can be justified by following qualitative reasons: in case of metastable event the trajectory scans one attractor and in $\lambda$-functional the conditional information on attractor's geometry $\omega$ is accumulated. In the coherent case, for arbitrary bounded "measurement time" $\Delta t$ (the time resolution is more than unit), in this measurement will take part a "superposition" of $n \approx \Delta t$ attractors-events. Differently, in the metastable case the measurement is a averaging on the outcomes belonging to single event. By contrast, in the coherent case the averaging occurs, in fact, on set of outcomes-events. In this definition it is important that the same event-attractor, in due course, can turn from metastable to coherent and \or on the contrary. Our technical identification attractor-event does not play a significant role here. This definition formalizes a change of dynamic modes, and it will be discussed in detail in the further.

3. The bifurcation relation expresses the duality between the environment and the trajectory in terms $\varpi$ and $\mu$. This circumstance is a reflection of the informational character of the interaction between them.

4. In the further we shall not consider the uninteresting, stationary situations: $\lambda_0 \cdot \log V < \varpi^{-1}$, is corresponding to delta-like density at limit $t \to \infty$. From a such stationarity it is possible to get rid, having used more fine partition or, speaking language of experimental physics, having increased resolution of the measurement tools. Note, that the trivial and pointwise partitions are result in the purely random dynamics.

## 4 Estimation of event's lifetime

**Definition 3**

Let at the moment $T$ the trajectory falls into a basin of attraction of some potential hole, and it has left them at the moment $T + \tau$. We shall name $\tau$ by *lifetime* of the corresponding metastable event-attractor.

In this sense, the frequency measure of attractor-event is:

$$\mu_i(T) = T^{-1} \sum_{r=1}^{R(T)} \tau_i(T_r),$$

$R(T)$-number of returns of the trajectory in the cell $\xi_i$ at the time $T$. $T_r$-time of $r$'s return. This is the relative, *total lifetime* of the event.
It is important, that $\tau = \tau(T)$ is time-dependent. This is an illustration of the time heterogeneity.
For us it is necessary, by virtue of (3), to estimate the time of evolution of the bifurcation parameter to magnitude $\varpi^{-1}$. We shall assume that attractor is contained in single cell: $\zeta_0$.
Let's define small parameter:

$$\varepsilon = \frac{t}{T} \ll 1,$$

$T$-time of $\zeta_0$ occurrence,

$t$-current time of this event existence (6)

Thus, global time is $T + t$. Here there are essentially various two cases:
1) The trajectory comes back in considered attractor-the reincarnations' cycle.

2) The trajectory gets in an "unknown vacuum", increasing a volume in definition (2).

Let's consider these cases separately.
We shall name the event $\xi_i$ *actual*, if $\varphi(T+t) \in \xi_i$.
For small evolution it is had:

$$\mu(T+t) = \frac{n(T)+t}{T+t} = \frac{\mu(T)+\varepsilon}{1+\varepsilon} \text{ -for actual event,}$$

$$\mu(T+t) = \frac{n(T)}{T+t} = \frac{\mu(T)}{1+\varepsilon} \qquad \text{-otherwise.}$$

Obtain:

$$\lambda(T+t) = \lambda_0 \cdot \left( \log V + \frac{\mu_0 + \varepsilon}{1+\varepsilon} \log\left(\frac{\mu_0+\varepsilon}{1+\varepsilon}\right) + \sum_{i=1}^{V-1} \frac{\mu_i}{1+\varepsilon} \log\left(\frac{\mu_i}{1+\varepsilon}\right) \right)$$

$$\lambda(T+t) > \varpi^{-1}.$$

It is easy to see, what solving of last inequality is transcendentally complex. Therefore, we shall receive square-law estimation. For this purpose we shall expand $\lambda$ in power series on small parameter $\varepsilon$ up to the second order. Necessity of the second order will be proved at calculations. It is also visible from Figure 1 and Figure 2.
We have:

$$\lambda(T+t) = \lambda(T,\varepsilon) = \lambda(T) + \lambda_\varepsilon \cdot \varepsilon + \lambda_{\varepsilon\varepsilon} \cdot \varepsilon^2 + o(\varepsilon^2).$$

After trivial calculations we obtain:

$$\lambda_\varepsilon = \lambda_0 \cdot \left( h(T) + \log\mu_0 \right)$$

$$\lambda_{\varepsilon\varepsilon} = \lambda_\varepsilon + \frac{\lambda_0}{2} \cdot \left( \mu_0^{-1} - 1 \right). \qquad (7)$$

2) In case of increasing volume $V \to V+1$ it is obtain:

$$\lambda(T+t) = \lambda(T) + \lambda_0 \cdot \log(1+V^{-1}) + \lambda_\varepsilon \cdot \varepsilon + \lambda_{\varepsilon\varepsilon} \cdot \varepsilon^2 + o(\varepsilon^2).$$

Here:

$$\mu_0 \approx T^{-1},$$
$$\lambda_\varepsilon \approx \lambda_0 \cdot (h(T) - \log T),$$
$$\lambda_{\varepsilon\varepsilon} \approx \lambda_\varepsilon + \frac{\lambda_0}{2} \cdot T \text{ for greater } T.$$

In Figure 3 it is visible (smoothed), that after occurrence of an "earlier unknown event" there is a "stress"-splash of information:

$$\delta\lambda \sim \lambda_0 \cdot V^{-1} \text{ caused by topological properties of an environment.}$$

However, then the entropy very quickly increases:

$$\partial_\varepsilon h \sim -\log\varepsilon, \varepsilon \approx T^{-1},$$

it leading to local relaxation (Figure 3). This stressful situation has two scenarios, depending on sign of quantity:

$$\lambda + \delta\lambda - \varpi^{-1} = \lambda + \lambda_0 \cdot \log(1+V^{-1}) - \varpi^{-1} \text{ and on "thickness" of}$$

attractor's walls. This is or a panic flight (the red schedule), or a fright, replaced by an intense interest (the green schedule). It is well visible in Figure 3. In the further, for estimations, we shall use the second scenario. Here we shall not estimate accuracy of approach. We are interested only in order of values and in qualitative picture.

We name the value $\lambda_\varepsilon$ by a *conditional, normalized speed of change of entropy*. Apparently, this quantity connects global $h$ and local $\log \mu$ information. Sign of this speed can be both positive, and negative. It explains two scenarios of behavior of the bifurcation parameter are represented in Figure 1 and Figure 2.

For the lifetime we obtain a rough estimate in linear approach:

$$\tau > \frac{\varpi^{-1} - \lambda(T)}{\lambda_\varepsilon} \cdot T = \frac{\varpi^{-1} - \lambda(T)}{\lambda_0 \cdot (h(T) + \log \mu_0)} \cdot T. \qquad (8)$$

For square-law approach it is necessary to find the solution of inequality:

$$\lambda(T) + \lambda_\varepsilon \cdot \varepsilon + \lambda_{\varepsilon\varepsilon} \cdot \varepsilon^2 > \varpi^{-1}.$$

We obtain:

$$\tau > \frac{-\lambda_\varepsilon + \sqrt{\lambda_\varepsilon^2 - 4\lambda_{\varepsilon\varepsilon}(\lambda - \varpi^{-1})}}{2\lambda_{\varepsilon\varepsilon}} \cdot T. \qquad (9)$$

Let's consider now two consecutive bifurcations: first, the trajectory falls into vacuum $\varpi_1$, then, after a while (lifetime), having experienced bifurcation, into a vacuum $\varpi_2$ ($\varpi_1 > \varpi_2$ -differently the trajectory will not observe $\varpi_2$ by virtue of the bifurcation relation (3)). As shown above it is possible to neglect a time of transition between attractors by virtue of it smallness in comparison with the lifetime. We shall accept also $h > -\log \mu_2$, as it is possible to use linear approximation for $\tau$ ( Figure 1), then:

$$\tau_2 \geq \left(\frac{1}{\varpi_2} - \frac{1}{\varpi_1}\right) \cdot \frac{T}{\lambda_0 \cdot \log V - \varpi_1^{-1} + \lambda_0 \cdot \log \mu_2} \qquad (10)$$

Or

$$\tau_2 \geq \frac{\Delta \omega}{\omega} \cdot \frac{T}{\lambda_0 \cdot \omega \cdot \log(V \cdot \mu_2) - 1}. \quad \text{for dense spectrum: } \frac{\Delta \omega}{\omega} \ll 1.$$

These estimations look like a uncertainty relation (non-standard, renormalized). It is possible to interpret $\varpi$ as a factor of energy. Really, locally in time (space) for an "eigenvalue" of the evolution operator it is obtain:

$$\partial_t \varphi \approx E \cdot \varphi,$$
$$E = E(t) = \lambda(t) \cdot \varpi.$$

For small time intervals- $\lambda \approx const$ and $E \sim \varpi$.

It justifies our interpretation. Thus, the lifetime can be characterized in terms of "energy jumps":

$$\tau = \tau(\varpi_i \to \varpi_j, T) = \tau(\varpi_i, \varpi_j, T) = \tau_{ij}(T).$$

## 5 Supermoduli–space. "Slow time"

Above, a local picture of the trajectory-environment interaction was considered. Now, we shall pass to a more global point of view.

Let's consider the environment as a set of connected attractors-vacua. For this purpose we define a quotient partition. It is easy to see that this is a topological, oriented graph:

$$G = (V, E).$$

Here we shall name this graph the *supermoduli–space* alike [1]. The attractors are corresponds to vertices vertexes $V$, to edges $E$-possible transitions between attractors. Each vertex is characterized by effective curvature (energy) of corresponding attractor-vacuum:

$$W : V \to \{\varpi\}.$$

Here $\{\varpi\}$-power spectrum of the environment potential-the set of attractors' energies (average curvatures for our estimations).

Two vertices vertex are connected by the edge:

$$e_{ij} \in E \subset V \times V,$$

if a direct transition between corresponding attractors is possible in the mechanism of bifurcation (3). The direction of an edge is defined by a direction of a trajectory transition. For each vertex $v \in V$ we shall designate a set of outgoing edges as $out(v)$. To each edge from this set we shall establish a correspondence with a transition probability between the vertices-attractors:

$$p : E \to (0,1), \sum_{j \in out(i)} p_{ij} = 1.$$

This randomness is caused by the noise of model-$\eta$. In future, for simplicity, we shall consider the lattice variant:

$p_{ij} = p_{ji} = (2 \dim M)^{-1}$, if $i, j$-are indexes of neighbouring cells, and

$p_{ij} = 0$ otherwise.

Actually, a distribution of transitive probabilities is generally non-uniform and it depends on local geometry of the environment potential graph (curvature tensor). For simplicity of further estimations we abstract from this information and accept spherically symmetric potential holes as attractors.

The vacua lifetimes give a natural partition of the time. This partition is not homogeneous, what is display a non-stationarity of our system. We take the factor of this partition and name this factor-time $\Theta$ as "slow time". In the spatial partition, accepted by us (2), it is a sequence is indexing the transitions between events-attractors. Thus, we obtain Markovian process $T$ on the graph $G$ i.e. the spatio-temporal factor.

**Definition 4**

We shall name triplet:

$$F \equiv (G, P, T) \tag{11}$$

the *factorsystem*.

By analogy to the theory of renewal processes $T$ is the operating (subordinate) Markovian process.

We have the factor-dynamics:

$$\ldots \to \xi_{i_\Theta} \to \xi_{i_{\Theta+1}} \to \ldots \to T^n \circ \xi_{i_\Theta} \to \ldots \tag{12}$$

and corresponding codynamics:

$$\ldots \to \varpi_{j_\Theta} \to \varpi_{j_{\Theta+1}} \to \ldots \to \varpi_{j_{\Theta+n}} \to \ldots \tag{13}$$

in the power spectrum of landscape.

Lifetime dynamics:

$$\ldots \to \tau_{k_\Theta} \to \tau_{k_{\Theta+1}} \to \ldots \to \tau_{k_{\Theta+n}} \to \ldots \tag{14}$$

Bifurcation relation (3) gives a quality to attractors (events). Therefore it is possible to define the "dynamics of quality":

$$\ldots \to \theta_{k_\Theta} \to \theta_{k_{\Theta+1}} \to \ldots \to \theta_{k_{\Theta+n}} \to \ldots \qquad (15)$$

Symbol $\theta_k$ reflects a metastable quality of attractor and has one of two values: metastable (singular), coherent.

## 6 Fluctuation, relaxation

It is almost obvious, that the $\lambda$-factor is a nonmonotonous function of time. That leads to destruction and occurrence of the attractors (15). By way of illustration we shall estimate the time of fluctuation and the time of relaxation and compare them.

Let's consider following sequence:

$$h \xrightarrow{t_F} (h - \Delta h) \xrightarrow{t_R} h,$$

$t_F$ -time of fluctuation,

$t_R$ -time of relaxation. $\qquad (16)$

The entropy fluctuation corresponds to fluctuation of density in the bifurcation mechanism. Therefore, as $t_F$ it is possible to accept the estimation for lifetime of metastable vacuum (8). It is the evolution of conditional entropy. Provided that the trajectory is localized in some partition's cell (it is trapped by metastable vacuum). This dynamic mode we shall name the *fluctuation mode*. The *relaxation*, in opposite, is an unconditional entropy growth. Such situation can arise, for example, in compact case: $V = const$ after a next "minimal bifurcation", when the relation (3) is true and curvature-energy of actual attractor is minimal. A long time after bifurcation (relaxation time) the trajectory does not notice others vacua and, thus, it is a sequence of coherent events. Thus, the system degenerates in an ordinary Markovian process on graph $G$, which actually coincides with the operating process (11).

Let's consider a compact case $V = const$. We assume, that in time $t_R$ the trajectory visits each cell of partition on average $t_R \cdot V^{-1}$ -times. It corresponds to a uniform, stationary distribution of the operating process (11). This is that we name *pre-ergodic consideration*.

In this case the entropy as a function of time has estimation:

$$h(T + t_F + t) \approx h(T + t_F) + h_\varepsilon \cdot \varepsilon,$$

$$h_\varepsilon = -(h + \overline{\log \mu}), \qquad (17)$$

$$\overline{\log \mu} = \frac{1}{V} \sum_i \log \mu_i (T + t_F) \text{ -pre-ergodic assumption,}$$

this estimation for the entropy production is analogously to (7),

$$\varepsilon = \frac{t_R}{T + t_F}.$$

According to (16) it is necessary to estimate $\varepsilon_R$ in the following equality:

$$\Delta h \approx -(h - \Delta h + \overline{\log \mu}) \cdot \varepsilon_R$$

Here
$$\Delta h \approx (h + \log \mu_0) \cdot \varepsilon_F \approx \lambda_0^{-1} \frac{\Delta \omega}{\omega^2},$$
that corresponds to the entropy fluctuation at transition $\omega_1 \to \omega_2, \omega_1 > \omega_2$
$$\Delta \omega = \inf_{\substack{\omega' \in \{\omega\} \\ \omega' > \omega}} |\omega - \omega'| \text{-minimal spectral gap.}$$

In the first infinitesimal order we obtain:
$$\frac{t_F}{t_R} \approx \frac{|\overline{\log \mu}| - h}{h + \log \mu_0} \approx \frac{\lambda_0 \omega^2 (|\overline{\log \mu}| - h)}{\Delta \omega}. \qquad (18)$$

For validity of linear approximation for a local entropy's evolution we assume what: $h + \log \mu_0 > 0$.

It is interesting to compare this value to unit.
$$\frac{t_F}{t_R} \sim 1, |\overline{\log \mu}| + |\log \mu_0| \approx 2h$$
$$\frac{t_F}{t_R} > 1, |\overline{\log \mu}| + |\log \mu| > 2h \qquad (19)$$

Certainly, it is a rough estimate. We see that ratio (19) is determined by location of the trajectory in the power spectrum, and by the power spectrum density. The dynamic mode of a fast diffusion exists in our system. It practically coincides with the operating process (11). Figuratively speaking, this mode allows the system to abstract from a "studied part of the graph", and concentrates its own "attention" and high transition properties: $\sqrt{D \cdot t} \sim \sqrt{(2 \cdot \dim M)^{-1} \cdot t_R}$ to a search of low-energy attractors.

From pre-ergodic reasons it is possible to replace, subgraph:
$$\forall V' \in V : \varpi(V') > \varpi_0$$
by one vertex with curvature (energy):
$$\varpi' = |V'|^{-1} \cdot \sum_{i \in V'} \varpi_i > \varpi_0,$$
that is for coherent event in sense dynamic factorization. This situation we shall name the *second order dynamic factorization*.

This example illustrates a presence of two dynamic modes in the system. The first, fluctuation is a localization of trajectory in the configuration space (moduli exploration). The second one is a relaxational–non local (supermoduli exploration). These modes replace each other in such manner that the fluctuations drift in a power spectrum of environment to energy minimum: to flatness.

We come to the important consequence.

**Proposition-definition**
> The Memory in the system: trajectory-environment (TE) leads to asymmetry of time that manifests itself as a quasi-directed drift of the metastable events in a direction of energy reduction-the *topological convergence*:

$$\varpi_{i_1} > \varpi_{i_2} > \ldots . \tag{20}$$

**Proof**

First, a sequence (20) is nonmonotonous, it is rather tendency image.

Let's consider the metastable event $\varpi_i > \varpi_{\min}$. A next metastable event can have both greater and smaller energy. We shall estimate the time $t^+$ of transition of the trajectory to the metastable event: $\varpi_j < \varpi_i$ and the time $t^-$ of transition to the metastable event: $\varpi_k > \varpi_i$. Then we shall compare them.

Let $T$ -moment of the bifurcation,

$$\varepsilon^- = \frac{t^-}{T},$$

$$\Delta_+\varpi = \inf_{k:\varpi_k>\varpi_i}(\varpi_k - \varpi_i) \text{ -top spectral gap,}$$

and let:

$$\Delta_+\varpi \ll \varpi_i.$$

Then in linear approach, using estimation for relaxation time (19), we obtain:

$$\varepsilon^- = \frac{t_R}{T} \approx -\frac{\Delta_+\varpi}{\omega^2} \cdot \frac{1}{\lambda_0(h+\log\mu)} > 0$$

This is the lower bound estimation for the return time $t^-$ of system in the power spectrum. It is easy to see, that for $t^+$ *such limitation does not exist*: As already next iteration can translate the system into the metastable phase. For this purpose it is necessary in order, the neighbourhood of the actual attractor (vacuum) contained some attractors (vacua), which have smaller energies (3).

More precisely, let

$V^- \equiv \{\forall v \in V : \omega(v) < \varpi_i\}$ -vacua set with energies smaller in comparison with the energy of actual vacuum. Then a probability that the trajectory will reach metastable event with *smaller energy* during the *relaxation time* is:

$$p_{t_R}(\omega_i \to \omega_j, \omega_j < \omega_i) \sim \sum_{t \leq t_R}\left(1 - \frac{V^-}{V}\right)^{t-1} \cdot \frac{V^-}{V} > 0.$$

In this estimation the Bernoulli tests instead of the Markovian process are used, that it is quite admissible at the most general assumptions of environment.

At the same time, as it was been shown, the probability of occurrence of metastable event with *greater energy* during this time ($t_R$) is *strong equal to zero* for a non-zero spectral gap. Figuratively, in order an interest to the history has arisen, it is necessary to partially forget it.

Here there is a phenomenon of the topological convergence; the convergence on an attractors' space (11), and in the environment-landscape spectra. Term "topological convergence" we have in sense of the dynamical factorization of the factor-system. This is a dynamical non-monotonous reduction of partition to trivial one. The fact, that the memory, being realized as a geometrical (metric) factor, leads to the topological convergence, is remarkable. It is possible to tell, that in our system the memory is a source of consciousness of the purpose.

The topological convergence is not conventional. Actually the system is nonstationary and the image of its evolution in the power spectrum of potential $\{\omega_i\}$ is not reduced to a simple convergence (regular, weak etc.). More likely, it reminds a turbulent flow. For illustration, even in case of the degenerated global minimum ($\{\omega_{\min,i} > 0\}, 0 < i < n, n > 1$) transitions between its non-connective components are possible. For this purpose, as it is easy to see, it is necessary that before a first passage time of a global minimum a condition:

$$\lambda_0 \cdot \log V > \omega_{\min}^{-1} \quad \text{(3) was satisfied.}$$

Continuing a parallel with quantum mechanics, it is possible to tell, that the system converges in "classical sense", as a sequence of decoherences. Even after a global minimum is achieved the interesting dynamics do not disappear. The Global minimum will collapse and a large relaxation may raise the trajectory in the landscape spectrum. And scenario will repeat: the large-scale oscillations.

## 7 Cycles on the Landscape spectrum. "Interference"

Here we shall quantitatively estimate the information sense in a phenomenon of the topological convergence. To this end we shall consider the cycles of operating process and their action in the real "fast" time.

Let:

$$\{V_i\} = \ldots \to V_{i_\Theta} \to V_{i_{\Theta+1}} \to \ldots \text{ trajectory of operating process on the}$$

supermoduli–space,
the corresponding sequence of vacua lifetimes:

$$\{\tau_i\} = \ldots \to \tau_{i_\Theta} \to \tau_{i_{\Theta+1}} \to \ldots$$

Unlike a parental process, $\tau_i$-is a strongly correlated. This is illustration of the information memory.

Scheme of our example is following: we shall consider a cycle of the operating process $\{V_i\}$ (11):

$$C(V_{i_0}, t_c) \equiv V_{i_0} \to V_{i_1} \to \ldots \to V_{i_N} \to V_{i_0}.$$

Let $t_c$ - a length of a cycle (in fast time)-

$$t_c = \sum_{k=0}^{N} \tau_{i_k}, \; \tau_{i_k} \text{-attractors' lifetimes.}$$

Any cycle is presented by the operator:

$$C^*(V_i, T, t_c) : \tau_i(T) \to \tau_i(T + t_c) \text{ (cocycle).}$$

The scenario can looks so: first, the trajectory, at the moment $T$ is grasped by attractor $\varpi_{i_0}$, there lives in it at some time (lifetime of metastable event $\tau_{i_0}(T)$), then the trajectory leaves the attractor and after time $t_c$ comes back again. During this time the system can or only relaxes, or it can tries the intermittent sequence of fluctuations and relaxations. We shall consider relaxation mode, as a most probable and simple. Clearly, that new event's lifetime $\tau_{i_0}(T + \tau_{i_0}(T) + t_c)$ will differ from $\tau_{i_0}(T)$ by virtue of time heterogeneity.

Moreover,
$$\exists t_0 : \forall t_c < t_0, \tau_{i_0}(T + \tau_{i_0} + t_c) < \tau_{i_0}(T).$$
We shall estimate the length of "minimal cycle", in this sense.

Figuratively being expressed, the trajectory is not stay too long there where was recently. This circumstance pays attention to the anomalous diffusion nature of considered system TE.

We start realization of this plan. Let's use a linear approach for lifetime estimation:
$$\Delta\lambda \approx \lambda_\varepsilon \cdot \varepsilon_c$$
$$\lambda_\varepsilon \approx \lambda_0(h + \overline{\log\mu}).$$

Here, as in (17) $\overline{\log\mu} \equiv |V|^{-1}\sum_i \log\mu_i$.

For simplicity of estimation, we implicitly assume that the piece of the trajectory corresponding to a "minimal cycle" will be uniformly distributed on all space of configurations. Obviously, it is not absolutely so. However, if we consider a cycle as a coherent (relaxation) phase, that is the most probable, our assumption is quite justified.

Obviously, $\lambda_\varepsilon \leq 0$.

We obtain for new, relative attractor's lifetime:
$$\varepsilon_2 = \frac{\tau_2}{T} \approx \frac{\Delta\lambda}{\lambda_0(\mathrm{T}^{t_c} \circ h + \log\mathrm{T}^{t_c} \circ \mu_0)},$$
$$\mathrm{T}^{t_c} \circ h \approx h + h_\varepsilon \cdot \varepsilon_c = h - (h + \overline{\log\mu})\varepsilon_c,$$
$$\log\mathrm{T}^{t_c} \circ \mu_0 \approx \log\mu_0(1+\varepsilon_c)^{-1}.$$

We must estimate $\varepsilon_c$ in equality:
$$\varepsilon_1 = \varepsilon_2.$$
More accurately:
$$\varepsilon_2 = \varepsilon_1(1 + \varepsilon_1 + \varepsilon_c).$$
Nevertheless, we interest the first order of smallness on $\varepsilon$. Therefore we shall consider the first variant.

Obtain:
$$\frac{-(h+\overline{\log\mu})\varepsilon_c}{h + (h+\overline{\log\mu})\varepsilon_c + \log\mu_0 - \log(1+\varepsilon_c)} \approx \varepsilon_1,$$
or, in first order of smallness:
$$\frac{\tau_c}{\tau_1} \approx \frac{\varepsilon_c}{\varepsilon_1} \approx -\frac{h + \log\mu_0}{h + \overline{\log\mu}}.$$

Here it is possible to note, that for the bifurcation moment: $h = \log V - (\lambda_0\omega_0)^{-1}$ (3). As one would expect, the obtained estimation does not differ from (19). We go to a conclusion that the time of correlation and the lifetime are values of same order. Their distinction is defined by non-uniformity of distribution of trajectory's points in configuration space. Quantitatively this heterogeneity can be expressed in terms: $\log\mu_0, \overline{\log\mu}$.

We have estimation for the period of the minimal cycle. Let's name this value the *correlation time*.

As it is easy to see, on a set of the vertexes-attractors visited in the past, the operating process can be considered as a sequence of the cycles closes. Such "interference of fluctuations" can leads to the dynamic factorization even of events with a least energy at present. Differently: frequent, fast trajectory returnings are destroying the metastable events translating them into coherent rank.

## 8 Memory and Dark energy

As we can see ([3](#)), after a determined level of decoherence, the blow-up like expansion is appear. Here we shall consider a speculative physical interpretation of this phenomenon in TE.

The expansion on configuration space is:

$$\frac{\Delta|\varphi|}{|\varphi|} \approx \frac{2 \cdot \lambda_\varepsilon \cdot \omega}{T} \cdot t \Delta t = H_\varphi(t) \cdot \Delta t,$$

$$H_\varphi = 2\lambda_\varepsilon \omega t T^{-1}.$$

Here

$T$ - whole time,

$t$ - time after performance of duality condition ([3](#)), see interior of navy rectangle on Figure [1](#),

$\omega$ - characteristic scale of a symmetry breaking,

$\lambda_\varepsilon = \lambda_0 \cdot (h + \log \mu)$ - is $\lambda$ product scale ([7](#)). For simplicity, we assume that this value do not significant change during lifetime (Figure [1](#)).

Factor $h + \log \mu$ has an information sense, which is a difference between global and local entropy. This "information defect" controls the accelerated expansion. Thus, in our toy model, the "dark energy" has information nature.

Differently, we may interpret the evolution of a single universe as a sequence of decoherences (progressive entanglement) ([ref](#)) that produces an effective environment $\omega(T)$ i.e. renormalization.

Note that $\lambda_\varepsilon \cdot T^{-1}$ is a dynamic quantity, which can have arbitrary small value. Its fine tuning happens by means of the previous history.

## 9 Discussion

Let us summarize what we have achieved so far. We have considered dynamics of system TE on various scales, and have obtained following qualitative picture:

The fluctuations of the trajectory's density distribution are alternated with relaxations. During fluctuation, the trajectory is localized in some vacuum and makes oscillations around of gradient flow equilibrium, absorbing the information on local geometry of an environment. Entropy, thus, eventually, decreases up to some threshold defined by the bifurcation condition ([3](#)). Further the trajectory has "blow-up like tunneling" into neighboring attractor-vacuum. After a fluctuation or a sequence of fluctuations ("soliton") the system is in a relaxation phase. The relaxation time depends on habitable volume, entropy and size of actual spectral gap (spectrum density).

First of all, we are interested in a fluctuation mode. After bifurcation, the metrics contains information on an environment, namely, about a local curvature of landscape potential-the symmetry breaking hierarchy.

We have shown that the image of dynamics in a landscape spectrum is regular in some informational (informational memory) sense. This large-scale topological convergence is important, non-obvious property, that generated by historically specified dynamics.

The measure, generated by trajectory, is non-singular and practically independent from initial conditions because of memory. The analog of dark energy (Hubble constant) in a configuration domain is non-constant and proportional to the difference between global and local landscape information. In renormalization domain this is an analog of the central charge that a measure of effective degrees of freedom. In TE the renormalization is a process with jumps between local theories. This seems like the renormgroup quantization.

Exploratory nature of TE is ideologically related to the "Census Bureau" concept [13]. The Memory in TE is a subject (explorer, not always having the Anthropic guise) of the multiverse-configuration space.

Renormalized interpretation of TE is justifying a very dense "discretuum" of the vacua energies [1]. It will denser in the future because of short "reincarnations' cycle".

## Acknowledgments

I am grateful to Alexei Morozov for a help on the publication of this paper.

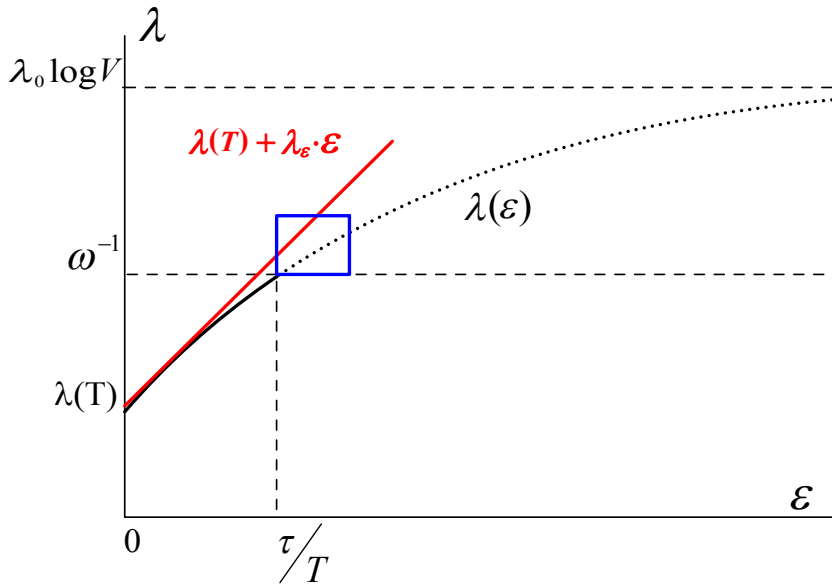

Figure 1: Linear approximation.

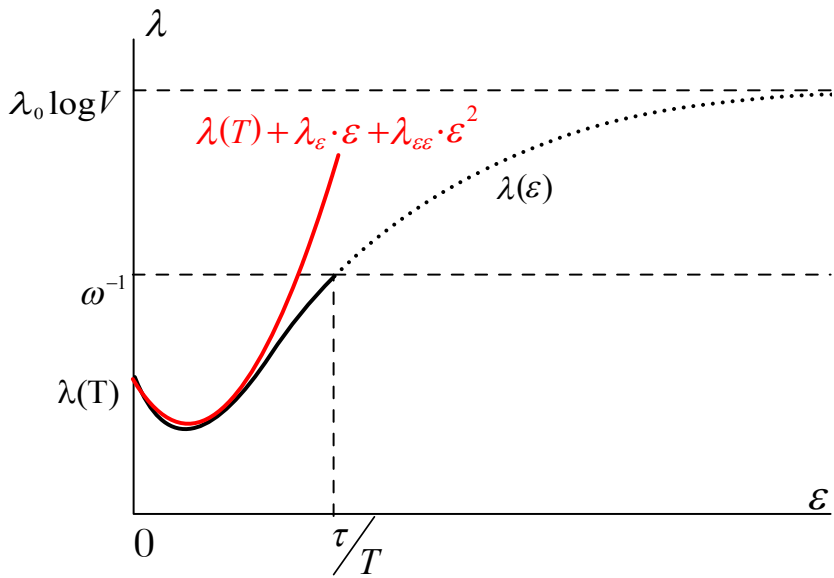

Figure 2: Quadratic approximation.

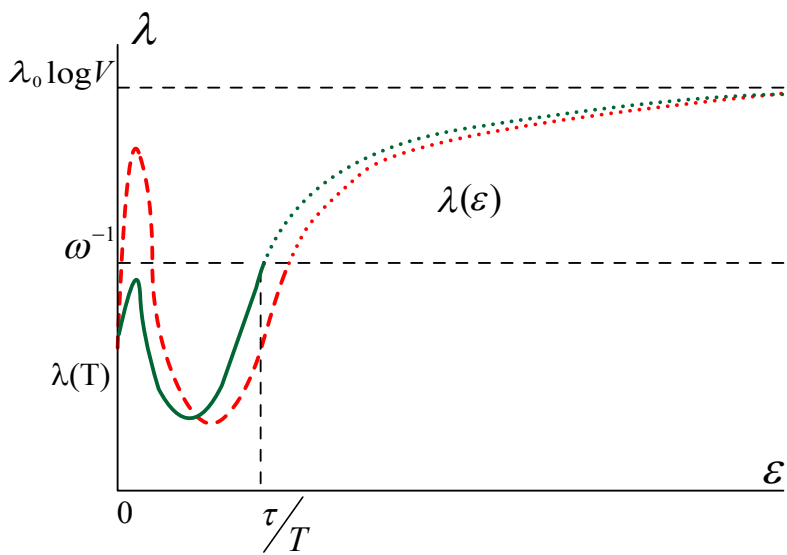

Figure 3: "Birth" of new event: $V \to V+1$, (smoothed).